\documentclass[a4paper]{article}

\usepackage{INTERSPEECH2021}
\usepackage{makecell}
\usepackage{booktabs}
\usepackage{bm}

\title{Detection of Lexical Stress Errors in Non-Native (L2) English with Data Augmentation and Attention}
\name{Daniel Korzekwa$^1$ $^2$, Roberto Barra-Chicote$^1$, Szymon Zaporowski$^2$, Grzegorz Beringer$^1$, Jaime Lorenzo-Trueba$^1$, Alicja Serafinowicz$^1$, Jasha Droppo$^1$, Thomas Drugman$^1$, Bozena Kostek$^2$}
\address{
$^1$Amazon TTS-Research \\
$^2$ Gdansk University of Technology, Faculty of ETI, Poland}
\email{korzekwa@amazon.com}

\begin{document}

\maketitle
\begin{abstract}

This paper describes two novel complementary techniques that improve the detection of lexical stress errors in non-native (L2) English speech: attention-based feature extraction and data augmentation based on Neural Text-To-Speech (TTS). In a classical approach, audio features are usually extracted from fixed regions of speech such as the syllable nucleus. We propose an attention-based deep learning model that automatically derives optimal syllable-level representation from frame-level and phoneme-level audio features. Training this model is challenging because of the limited amount of incorrect stress patterns. To solve this problem, we propose to augment the training set with incorrectly stressed words generated with Neural TTS. Combining both techniques achieves 94.8\% precision and 49.2\% recall for the detection of incorrectly stressed words in L2 English speech of Slavic and Baltic speakers.

\end{abstract}

\section{Introduction}
\label{sec:intro}

Computer Assisted Pronunciation Training (CAPT) usually focuses on practicing pronunciation of phonemes \cite{witt2000phone, leung2019cnn, korzekwa2020uncertainty}, while there is evidence in non-native (L2) English speakers that practicing lexical stress improves speech intelligibility \cite{field2005intelligibility,lepage2014intelligibility}. Lexical stress is a syllable-level phonological feature. It is a part of the phonological rules that define how words should be spoken in a given language. Stressed syllables are usually longer, louder, and expressed with a higher pitch than their unstressed counterparts \cite{jung2018acoustic}. Lexical stress is inter-connected with phonemic representation. For example, placing lexical stress on a different syllable of a word may lead to different phonemic realizations known as `vowel reduction' \cite{bergem1991acoustic}.

The focal point of our work is the detection of words with incorrect stress patterns. The training data with human speech is usually highly imbalanced, with few training examples of incorrectly stressed words. It makes training machine learning models for this task challenging. We address this problem by augmenting the training set with synthetic speech that is generated with Neural Text-To-Speech (TTS) \cite{latorre2019effect}. Neural TTS allows us generating words with both correct and incorrect stress patterns.

Most of the existing approaches for automated lexical stress assessment are based on carefully designed features that are extracted from fixed regions of speech signal such as the syllable nucleus \cite{ferrer2015classification,shahin2016automatic, chen2010automatic_2}. We introduce attention mechanism \cite{vaswani2017attention} to automatically learn optimal syllable-level representation. 
Attention-based approach originates from the intuition of how people detect specific patterns in high dimensional and unstructured data such as visual and speech signals \cite{posner1990attention}. For example, we might focus our attention on the duration ratio between nuclei of two neighboring syllables, incidentally, an important predictor of lexical stress. The syllable-level representation is derived from frame-level (F0, intensity) and phoneme-level (duration) audio features and the corresponding phonetic representation of a word. We do not indicate precisely the regions of the audio signal that are important for the detection of lexical stress errors. The attention mechanism does it automatically.

To the best of our knowledge, this paper is the first attempt, for the task of lexical stress error detection, to: \emph{i)} augment the training data with Neural TTS, \emph{ii)} use attention mechanisms to automatically extract syllable-level features for lexical stress error detection. Ruan et al. \cite{ruan2019end} used attention-based architecture of transformers for lexical stress detection. However, their paper concerns recognizing stressed and unstressed phonemes. They do not detect lexical stress errors, which is crucial in CAPT applications.

The paper is structured as follows. In Section \ref{sec:related_work}, we review the related work. Section \ref{sec:proposed_model} describes the proposed model. Section \ref{sec:speech_corpora} reviews human and synthetic speech corpora. In Section \ref{sec:experiments}, we present our experiments, and Section \ref{sec:conclusion} concludes the paper.

\section{Related Work}
\label{sec:related_work}

\begin{figure*}[!t]
  \centering
   \caption{Attention-based Deep Learning model for the detection of lexical stress errors.}
  \includegraphics[height=4.0cm]{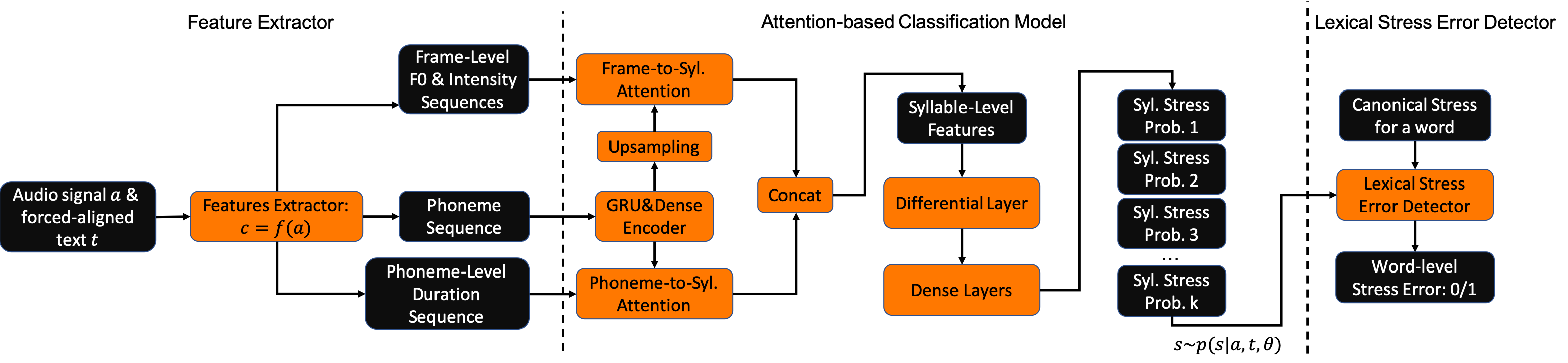}
  \label{fig:model_architecture}
\end{figure*}

The existing work focuses on the supervised classification of lexical stress using Neural Networks \cite{li2018automatic, shahin2016automatic}, Support Vector Machines \cite{chen2010automatic_2,zhao2011automatic} and Fisher’s linear
discriminant \cite{chen2007using}. There are two popular variants: a) discriminating syllables between primary stress/no stress \cite{ferrer2015classification}, and b) classifying between primary stress/secondary stress/no stress \cite{li2013lexical,li2018automatic}. Ramanathi et al. \cite{ramanathi2019asr} have followed an alternative unsupervised way of classifying lexical stress, which is based on computing the likelihood of an acoustic signal for a number of possible lexical stress representations of a word.

Accuracy is the most commonly used performance metric, and it indicates the ratio of correctly classified stress patterns on a syllable \cite{li2013lexical} or word level \cite{chen2010automatic_2}. On the contrary, following Ferrer et al. \cite{ferrer2015classification}, we analyze precision and recall metrics because we aim to detect lexical stress errors and not just classify them.

Existing approaches for the classification and detection of lexical stress errors are based on carefully designed features. They start with aligning a speech signal with phonetic transcription, performed via forced-alignment \cite{shahin2016automatic, chen2010automatic_2}. Alternatively, Automatic Speech Recognition (ASR) can provide both phonetic transcription and its alignment with a speech signal \cite{li2013lexical}. Then, prosodic features such as duration, energy and pitch \cite{chen2010automatic_2} and cepstral features such as MFCC and Mel-Spectrogram \cite{ferrer2015classification,shahin2016automatic} are extracted. These features can be extracted on the syllable \cite{shahin2016automatic} or syllable nucleus \cite{ferrer2015classification,chen2010automatic_2} level. 

Shahin et al. \cite{shahin2016automatic} computed features of neighboring vowels, and Li et al. \cite{li2013lexical} included the features for two preceding and two following syllables in the model. The features are often preprocessed and normalized to avoid potential confounding variables \cite{ferrer2015classification}, and to achieve better model generalization by normalizing the duration and pitch on a word level \cite{ferrer2015classification,chen2007using}. Li et al. \cite{li2018automatic} added canonical lexical stress to input features, which improves the accuracy of the model. 

In our approach, we use attention mechanisms to derive automatically regions of the audio signal that are important for the detection of lexical stress errors. We also use data augmentation through the generation of artificial data with Neural TTS.

\section{Proposed Model}
\label{sec:proposed_model}

The proposed model consists of three subsystems: Feature Extractor, Attention-based Classification Model, and Lexical Stress Error Detector. It is illustrated in Figure \ref{fig:model_architecture}.

\subsection{Feature Extractor}
\label{ssec:feature_extractor}

The Feature Extractor extracts prosodic features and phonemes from speech signal $\mathbf{a}$ and forced-aligned text $\mathbf{t}$. To obtain forced-alignment, we used Montreal toolkit \cite{mcauliffe2017montreal} along with an acoustic model pretrained on LibriSpeech ASR corpus \cite{panayotov2015librispeech}. The prosodic features $\mathbf{c}=f(\mathbf{a})$ are formed by: F0, intensity [dB SPL] and phoneme-level durations. The F0 and intensity features are computed at the frame level using Praat library \cite{boersma2006praat} (time step: 10 ms, window size: 40 ms). The F0 contour is linearly interpolated in unvoiced regions. These raw features will be further transformed by the attention-based model to the syllable-level representation.

\subsection{Attention-based Classification Model}
\label{ssec:classification_model}

The Attention-based Classification Model maps frame-level and phoneme-level features to the syllable-level representation. Then, it produces a lexical stress pattern $\mathbf{s}$, modeled as a sequence of Bernoulli random variables $\mathbf{s}=\{s_1,..,s_k\}$ (stressed/unstressed) over $K$ syllables of a multi-syllable word, conditioned on audio $\mathbf{a}$ and text $\mathbf{t}$ representations. Let us define it as a conditional probability distribution $\mathbf{s} \sim p(\mathbf{s}|\mathbf{a},\mathbf{t},\bm{\theta})$, where $\bm{\theta}$ are the parameters of the model.

\begin{figure*}[!t]
  \centering
   \caption{Top: forced-alignment mapping between phonemes and frames for the word 'garage'. Middle: Frame-to-syllable attention weights matrix. Bottom: (Sub)Phoneme-to-syllable attention weights matrix.}
  \label{fig:frame_phoneme_level_att}
  \includegraphics[width=11cm]{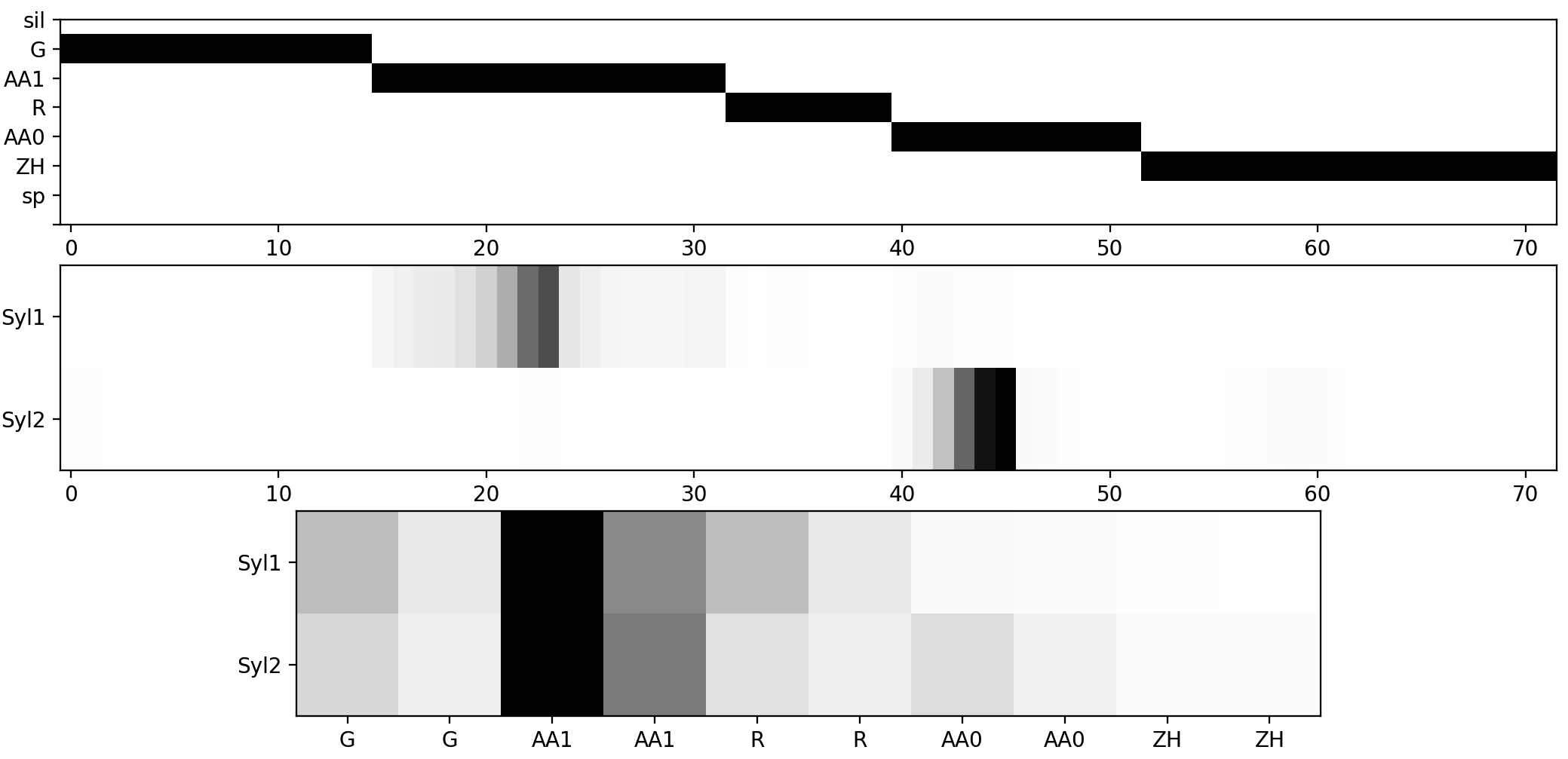}
\end{figure*}

To extract syllable-level features, we use two dot-product attentions operating on the frame and phoneme levels. To build better intuition on what these two attention do, in Figure \ref{fig:frame_phoneme_level_att} we show the frame-level and phoneme-level attention plots for the word 'garage' pronounced by a Polish speaker and incorrectly stressed on the first syllable in reference to American English. This word has a similar pronunciation but different lexical stress in Polish and American English languages (`G AA1 R AA0 ZH' vs `G ER0 AA1 ZH'). Both attentions find the most relevant regions of the frame-level and phoneme-level features.

The dot-product attention is presented in Equation \ref{eq_att}, and it follows the notation proposed by Vaswani et al. \cite{vaswani2017attention}. It is based on three inputs: Query ($\mathbf{Q}$), Keys ($\mathbf{K}$) and Values ($\mathbf{V}$), where $d_k$ is the dimensionality of $\mathbf{K}$.

\begin{equation}
Attention(\mathbf{Q}, \mathbf{K}, \mathbf{V}) = softmax(\dfrac{\mathbf{Q}\mathbf{K}^t}{\sqrt{d_k}})\mathbf{V}
\label{eq_att}
\end{equation}

The attention inputs are represented as follows.  Query refers to the syllable positional embeddings defined by one-hot syllable index encodings. Keys represents a sequence of sub-phonemes. Each sub-phoneme is represented by a set of features: $phoneme\_id$, $syllable\_index$, $is\_vowel$, $left\_or\_right\_sub\_phoneme$. All features are one-hot encoded and processed with a Gated Recurrent Unit (GRU) layer \cite{cho2014learning} (units:4, dropout: 0.24). In the end, encoded sub-phoneme sequence is passed through linear dense layers. In the case of the frame-level attention, the encoded sub-phoneme sequence is upsampled to the frame level using phoneme durations from forced-alignment. In upsampling, we simply replicate phonemes across aligned frames of audio signal. Similar phoneme-to-frame upsampling has been recently adopted in Text-To-Speech \cite{elias2020parallel}. Finally, Values are the $F0/intensity$ and $duration$ features for frame-level and phoneme-level attentions respectively.

To model relative prominence, we introduce a differential bi-directional layer that computes the ratios of syllable-level acoustic features for each syllable and its two neighbors (Figure \ref{fig:model_architecture}). The bi-directional layer is implemented as a simple `division' math operation and it does not contain any trainable parameters. The output of the differential layer is further processed by three dense layers (units: 4, activation: tanh, dropout: 0.24), followed by a linear dense layer (units: 2, dropout: 0.24) that produces a two-dimensional output for each syllable. It is then squeezed by a softmax function to generate lexical stress probabilities.

\subsection{Training of the Classification Model}
\label{ssec:training_procedure}

We train the model on a set of $N$ triplets that contains 1) human recorded words and 2) synthetic words generated using Neural TTS. A single triplet is represented by $\{\mathbf{s_n},\mathbf{a_n},\mathbf{t_n}\}$, where $n=1..N$ is the index of a training example.

The concept of data augmentation can be explained using a framework of Bayesian Inference. Consider three random variables, lexical stress $\mathbf{s_n}$, audio signal $\mathbf{a_n}$ and text $\mathbf{t_n}$. All variables are observed for the training examples of human speech. However, for the synthetic speech, we only observe the lexical stress and text variables. The audio signal is unobserved (hidden) because we have to generate it.

To train this model, we derive a negative log-likelihood loss over a joint probability distribution of lexical stress $\mathbf{s}$ and audio $\mathbf{a}$ random variables, as depicted in Equation \ref{eq1}. The loss is further approximated with the variational lower bound \cite{jordan1999introduction}, as presented in Equation \ref{eq2} (we omit $\bm{\theta}$ for brevity). For the training examples of synthetic speech, the conditional probability distribution over the audio signal $\mathbf{a_n} \sim p(\mathbf{a_n}|\mathbf{s_n},\mathbf{t_n})$ is estimated with Neural TTS, and for human recorded words, it is given explicitly.

\begin{equation}
\mathcal{L(\bm{\theta})} = -\sum_n^N log \int p(\mathbf{s_n},\mathbf{a_n}|\mathbf{t_n},\bm{\theta})d\mathbf{a_n}
  \label{eq1}
\end{equation}

\begin{equation}
 log \int p(\mathbf{s_n},\mathbf{a_n}|\mathbf{t_n})\mathbf{da_n}\approx E_{\mathbf{a_n}\sim p(\mathbf{a_n}|\mathbf{t_n},\mathbf{s_n})}[logp(\mathbf{s_n}|\mathbf{a_n},\mathbf{t_n})]
  \label{eq2}
\end{equation}

The model was implemented in MxNet \cite{chen2015mxnet}, trained with Stochastic Gradient Descent optimizer (learning rate: 0.1, batch size: 20) and tuned with Bayesian optimization \cite{paleyes2019emulation}. Training data were split into buckets based on the number of frames in an audio signal, using Gluon-NLP package \cite{guo2020gluoncv}. A single bucket contains words with the same number of syllables with zero-padded acoustic and sub-phoneme sequences.

\subsection{Lexical Stress Error Detector}
\label{ssec:error_detector}

The Lexical Stress Error Detector reports on lexical stress error if the expected (canonical) and estimated lexical stress for a given syllable do not match and the corresponding probability is higher than a given threshold.

\section{Speech Corpus}
\label{sec:speech_corpora}

Our speech corpus consists of human and synthetic speech. The data were split into training and testing sets with disjointed speakers ascribed to each set. Human speech contains L1 and L2 speakers of English. Synthetic data were generated with Neural TTS and are included only in the training set. All audio files were downsampled to a 16 kHz sampling rate. The data are summarized in Table \ref{tab:speech_corpora_traintest}, and we provide more details in the following subsections.

\begin{table}[!ht]
\scriptsize
  \caption{Train and test sets details.}
  \label{tab:speech_corpora_traintest}
  \centering
  \begin{tabular}{llll}
    \toprule  
    \textbf{Data set}  & \textbf{\makecell[l]{Speakers \\ (L2)}} & \textbf{\makecell[l]{Words \\ (unique)}} &  \textbf{\makecell[l]{Stress \\ Errors}} \\
    \midrule
    Train set (human) & 473 (10) & 8223 (1528) & 425\\
    Train set (TTS) & 1 (0) & 3937 (1983) & 2005\\
    Test set (human) & 176 (21) & 2108 (378) & 189\\
    \bottomrule
  \end{tabular}

\end{table}

\subsection{Human Speech}
\label{ssec:human_speech}

Due to the limited availability of L2 corpora, we recorded our own L2-English corpus of Slavic and Baltic speakers. It also allows us to evaluate the model during interactive English learning sessions with our students. The corpus contains speech from 25 speakers (23 Polish, 1 Ukrainian and 1 Lithuanian): 7 females and 18 males, all between 24 and 40 years old. All speakers read a list of two hundred words. One hundred words were prepared by a professional English teacher, including frequently mispronounced words by Slavic and Baltic students. The second half consists of the most common words that were obtained from Google's Trillion Word Corpus \cite{michel2011quantitative} based on n-gram frequency analysis. We excluded abbreviations and one-syllable words.

Additionally, L1 and L2 English speech was collected from publicly available speech data sets, including TIMIT \cite{garofolo1993darpa}, Arctic \cite{kominek2004cmu}, L2-Arctic \cite{zhao2018l2} and Porzuczek \cite{porzuczek2017english}. 

\subsection{Synthetic Speech}
\label{ssec:synthetic_speech}

Complementary to human recordings, synthetic speech was generated with Neural TTS by Latorre et al. \cite{latorre2019effect}. The Neural TTS consists of two modules. Context-generation module is an attention-based encoder-decoder neural network that generates a mel-spectrogram from a sequence of phonemes. Then, a Neural Vocoder converts it to the speech signal. The Neural Vocoder is a neural network of architecture similar to the work by \cite{oord2018parallel}. The Neural TTS was trained using speech of a professional American voice talent. To generate words with different lexical stress patterns, we modify lexical stress markers associated with the vowels in the phonemic transcription of a word. For example, with the input of /r iy1 m ay0 n d/ we can place lexical stress on the first syllable of the word `remind'. 1980 popular English words were synthesized with correct and incorrect stress patterns.

\subsection{Lexical Stress Annotations}
\label{ssec:lexical_stress_annotations}

L1 corpora were segmented into words and annotated automatically using a proprietary Amazon American English Lexicon, taking into account the syntactic context of the word. Neural TTS speech and the speech of L2 speakers were annotated by 5 American English linguists into `primary' and `no stress' categories, keeping the words for which a minimum of 4 out of 5 linguists agreed on the stress pattern. Annotators were not able to distinguish between primary and secondary lexical stress.
81.5\% of synthesized words matched the intended stress patterns with a minimum of 4 annotators' agreement. It shows that Neural TTS can be used to generate incorrectly stressed speech.

\section{Experiments}
\label{sec:experiments}

\begin{figure}[!t]
  \centering
  \caption{Precision-recall curves for evaluated systems.}
  \includegraphics[width=7.4cm]{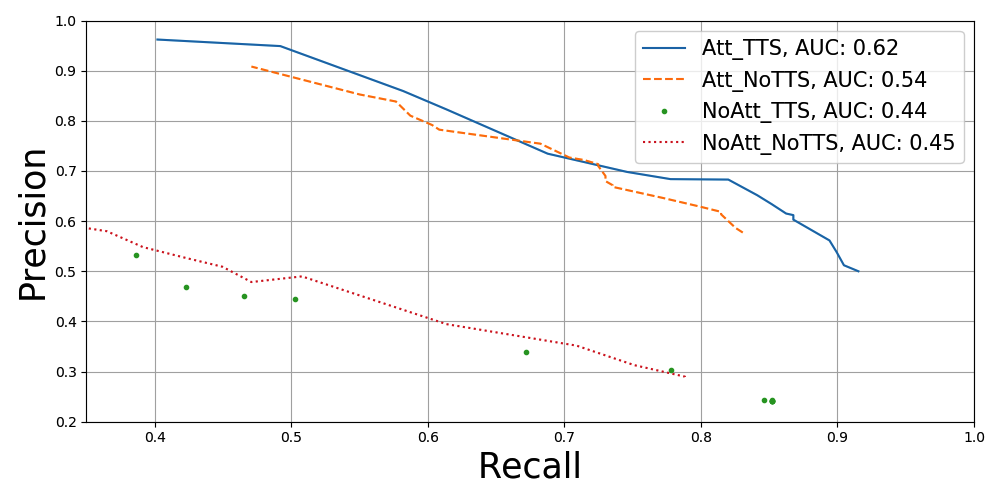}
  \label{fig:precision_recall_plot}
\end{figure}

The proposed model (Att\_TTS) from Section \ref{sec:proposed_model}  is compared to three baseline models that are designed to measure the impact of the Neural TTS data augmentation and the attention mechanism. To compare these models, we plotted their precision-recall curves and gave their corresponding area under a curve (AUC) along with our results, see Figure \ref{fig:precision_recall_plot}.

The Att\_NoTTS model has the same architecture as the Att\_TTS, but the synthetic speech is excluded from the `training set'. The NoAtt\_TTS model uses the same training set as the Att\_TTS, but it has no attention mechanism. Instead, as a syllable-level representation, it uses mean values of acoustic features for the corresponding syllable nucleus. The NoAtt\_NoTTS model has no attention, and it does not use Neural TTS data augmentation.

As a state-of-the-art baseline, we use the work by Ferrer et al. \cite{ferrer2015classification}. However, a direct comparison is not possible. In their test corpus, there were 46.4\% (191 out of 411) of incorrectly stressed words, far more than 9.4\% (189 out of 2109) words in our experiment. The fewer lexical stress errors are made by users, the more challenging it is to detect it. They also used proprietary L2 English of Japanese speakers. Due to the lack of available benchmark and standard speech corpora for the task of lexical stress assessment, we could not make a fairer comparison with the state-of-the-art.

\label{ssec:experiments_results}

\subsection{Experimental Results}

First, we compare Att\_NoTTS and NoAtt\_NoTTS models. Using the attention mechanism for automatic extraction of syllable-level features significantly improves the detection of lexical stress errors. It is illustrated by precision-recall curves and AUC metric in Figure \ref{fig:precision_recall_plot}. To be comparable with the study by Ferrer et al., we fix recall to around 50\% and compare the models using precision as shown in Table \ref{tab:precision_recall_acc}.

The Att\_NoTTS attention-based can be further improved. Augmenting the training set with incorrectly stressed words (Att\_TTS) boosts precision from 87.85\% to 94.8\%, at a recall level of 50\%. Data augmentation helps because it increases the number of words with incorrect stress patterns in the training set. It prevents the model from exploiting a strong correlation between phonemes and lexical stress in correctly stressed words. Using data augmentation in the simpler no-attention-based model (NoAtt\_TTS) does not help. It is because NoAtt\_TTS uses only prosodic features for fixed regions of speech, so this model cannot overfit to phonetic input.

\begin{table}[!ht]
\scriptsize
 \caption{Precision and recall [\%, 95\% Confidence Interval] of detecting lexical stress errors, at around 50\% recall. * - Ferrer et al. model has been evaluated on the data with 46.4\% of lexical stress errors, compared to 9.4\% of errors on our data set. This data point indicates that our proposed model AttTTS should outperform Ferrr et al. model if both were evaluated exactly in the same conditions.}
  \label{tab:precision_recall_acc}
  \centering
  \begin{tabular}{lll}
    \toprule  
   Model & Precision & Recall  \\
    \midrule
    AttTTS & 94.8 (89.18-98.03) & 49.2 (42.13-56.3) \\
    AttNoTTS & 87.85 (80.67-93.02) & 49.74 (42.66-56.82)\\
    NoAttTTS & 44.39 (37.85-51.09) & 50.26 (43.18-57.34)  \\
    NoAttNoTTS & 48.98 (42.04-55.95) & 50.79 (43.70-57.86)  \\
    Ferrer et al. \cite{ferrer2015classification} * & 95.00 (na-na) & 48.3 (na-na)  \\
    \bottomrule
  \end{tabular}
\end{table}

Ferrer et al. \cite{ferrer2015classification} reported on a similar performance to our Att\_TTS model with a precision of 95\% and a recall of 48.3\% on L2 English speech of Japanese speakers. However, in their testing data, the proportion of incorrectly stressed words is much larger, which makes it easier to detect lexical stress errors.

\section{Conclusion and Future Work}
\label{sec:conclusion}

Using an attention-based neural network for the automatic extraction of syllable-level features significantly improves the detection of lexical stress errors in L2 English speech, compared to baseline models. However, this model has a tendency to classify lexical stress based on highly-correlated phonemes. We can counteract this effect by augmenting the training set with incorrectly stressed words generated with Neural TTS. It boosts the performance of the attention-based model by 14.8\% in the AUC metric and by 7.9\% in precision, while maintaining recall at a level close to 50\%. Data Augmentation, however, does not help when applied to a simpler model without an attention mechanism.

We found that the current word-level model is not able to correctly classify lexical stress when two words are linked \cite{hieke1984linking} and stress shift may occur  \cite{shattuck1994stress}. For example, two neighboring phonemes /er/ in the text `her arrange' /hh er - er ey n jh/ are pronounced as a single phoneme. Therefore, in future, we plan to move away from the assessment of isolated words and extend the current model to detect lexical stress errors at the sentence level. 
We plan to replace a single-speaker TTS model to generate synthetic lexical stress errors with a multi-speaker model. We plan to analyze the accuracy of detecting lexical stress errors for speakers with different proficiency levels of English.

\bibliographystyle{IEEEtran}

\bibliography{lexical_stress_is2021}

\begin{thebibliography}{10}
\providecommand{\url}[1]{#1}
\csname url@samestyle\endcsname
\providecommand{\newblock}{\relax}
\providecommand{\bibinfo}[2]{#2}
\providecommand{\BIBentrySTDinterwordspacing}{\spaceskip=0pt\relax}
\providecommand{\BIBentryALTinterwordstretchfactor}{4}
\providecommand{\BIBentryALTinterwordspacing}{\spaceskip=\fontdimen2\font plus
\BIBentryALTinterwordstretchfactor\fontdimen3\font minus
  \fontdimen4\font\relax}
\providecommand{\BIBforeignlanguage}[2]{{%
\expandafter\ifx\csname l@#1\endcsname\relax
\typeout{** WARNING: IEEEtran.bst: No hyphenation pattern has been}%
\typeout{** loaded for the language `#1'. Using the pattern for}%
\typeout{** the default language instead.}%
\else
\language=\csname l@#1\endcsname
\fi
#2}}
\providecommand{\BIBdecl}{\relax}
\BIBdecl

\bibitem{witt2000phone}
S.~M. Witt and S.~J. Young, ``Phone-level pronunciation scoring and assessment
  for interactive language learning,'' \emph{Speech communication}, vol.~30,
  no. 2-3, pp. 95--108, 2000.

\bibitem{leung2019cnn}
W.-K. Leung, X.~Liu, and H.~Meng, ``Cnn-rnn-ctc based end-to-end
  mispronunciation detection and diagnosis,'' in \emph{ICASSP 2019-2019 IEEE
  International Conference on Acoustics, Speech and Signal Processing
  (ICASSP)}.\hskip 1em plus 0.5em minus 0.4em\relax IEEE, 2019, pp. 8132--8136.

\bibitem{korzekwa2020uncertainty}
D.~Korzekwa, J.~Lorenzo-Trueba, S.~Zaporowski, S.~Calamaro, T.~Drugman, and
  B.~Kostek, ``Mispronunciation detection in non-native (l2) english with
  uncertainty modeling,'' in \emph{Accepted to ICASSP 2021 IEEE International
  Conference on Acoustics, Speech and Signal Processing (ICASSP)}.\hskip 1em
  plus 0.5em minus 0.4em\relax IEEE, 2021.

\bibitem{field2005intelligibility}
J.~Field, ``Intelligibility and the listener: The role of lexical stress,''
  \emph{TESOL quarterly}, vol.~39, no.~3, pp. 399--423, 2005.

\bibitem{lepage2014intelligibility}
A.~Lepage and M.~G. Bus{\`a}, ``Intelligibility of english l2: The effects of
  incorrect word stress placement and incorrect vowel reduction in the speech
  of french and italian learners of english,'' in \emph{Proc. of the Intl.
  Symposium on the Acquisition of Second Language Speech}, vol.~5, no. 2014,
  2014, pp. 387--400.

\bibitem{jung2018acoustic}
Y.-J. Jung, S.-C. Rhee \emph{et~al.}, ``Acoustic analysis of english lexical
  stress produced by korean, japanese and taiwanese-chinese speakers,''
  \emph{Phonetics and Speech Sciences}, vol.~10, no.~1, pp. 15--22, 2018.

\bibitem{bergem1991acoustic}
D.~R.~v. Bergem, ``Acoustic and lexical vowel reduction,'' in \emph{Phonetics
  and Phonology of Speaking Styles}, 1991.

\bibitem{latorre2019effect}
J.~Latorre, J.~Lachowicz, J.~Lorenzo-Trueba, T.~Merritt, T.~Drugman,
  S.~Ronanki, and V.~Klimkov, ``Effect of data reduction on
  sequence-to-sequence neural tts,'' in \emph{ICASSP 2019-2019 IEEE
  International Conference on Acoustics, Speech and Signal Processing
  (ICASSP)}.\hskip 1em plus 0.5em minus 0.4em\relax IEEE, 2019, pp. 7075--7079.

\bibitem{ferrer2015classification}
L.~Ferrer, H.~Bratt, C.~Richey, H.~Franco, V.~Abrash, and K.~Precoda,
  ``Classification of lexical stress using spectral and prosodic features for
  computer-assisted language learning systems,'' \emph{Speech Communication},
  vol.~69, pp. 31--45, 2015.

\bibitem{shahin2016automatic}
M.~A. Shahin, J.~Epps, and B.~Ahmed, ``Automatic classification of lexical
  stress in english and arabic languages using deep learning.'' in
  \emph{INTERSPEECH}, 2016, pp. 175--179.

\bibitem{chen2010automatic_2}
J.-Y. Chen and L.~Wang, ``Automatic lexical stress detection for chinese
  learners' of english,'' in \emph{2010 7th Intl. Symposium on Chinese Spoken
  Language Processing}.\hskip 1em plus 0.5em minus 0.4em\relax IEEE, 2010, pp.
  407--411.

\bibitem{vaswani2017attention}
A.~Vaswani, N.~Shazeer, N.~Parmar, J.~Uszkoreit, L.~Jones, A.~N. Gomez,
  L.~Kaiser, and I.~Polosukhin, ``Attention is all you need,'' in
  \emph{Advances in neural information processing systems}, 2017, pp.
  5998--6008.

\bibitem{posner1990attention}
M.~I. Posner and S.~E. Petersen, ``The attention system of the human brain,''
  \emph{Annual review of neuroscience}, vol.~13, no.~1, pp. 25--42, 1990.

\bibitem{ruan2019end}
Y.~Ruan, X.~Wang, H.~Liu, Z.~Ou, Y.~Gao, J.~Cheng, and Y.~Qian, ``An end-to-end
  approach for lexical stress detection based on transformer,'' \emph{arXiv
  preprint arXiv:1911.04862}, 2019.

\bibitem{li2018automatic}
K.~Li, S.~Mao, X.~Li, Z.~Wu, and H.~Meng, ``Automatic lexical stress and pitch
  accent detection for l2 english speech using multi-distribution deep neural
  networks,'' \emph{Speech Communication}, vol.~96, pp. 28--36, 2018.

\bibitem{zhao2011automatic}
J.~Zhao, H.~Yuan, J.~Liu, and S.~Xia, ``Automatic lexical stress detection
  using acoustic features for computer assisted language learning,''
  \emph{Proc. APSIPA ASC}, pp. 247--251, 2011.

\bibitem{chen2007using}
N.~Chen and Q.~He, ``Using nonlinear features in automatic english lexical
  stress detection,'' in \emph{2007 Intl. Conference on Computational
  Intelligence and Security Workshops (CISW 2007)}.\hskip 1em plus 0.5em minus
  0.4em\relax IEEE, 2007, pp. 328--332.

\bibitem{li2013lexical}
K.~Li, X.~Qian, S.~Kang, and H.~Meng, ``Lexical stress detection for l2 english
  speech using deep belief networks.'' in \emph{Interspeech}, 2013, pp.
  1811--1815.

\bibitem{ramanathi2019asr}
M.~K. Ramanathi, C.~Yarra, and P.~K. Ghosh, ``Asr inspired syllable stress
  detection for pronunciation evaluation without using a supervised classifier
  and syllable level features.'' in \emph{INTERSPEECH}, 2019, pp. 924--928.

\bibitem{mcauliffe2017montreal}
M.~McAuliffe, M.~Socolof, S.~Mihuc, M.~Wagner, and M.~Sonderegger, ``Montreal
  forced aligner: Trainable text-speech alignment using kaldi.'' in
  \emph{Interspeech}, vol. 2017, 2017, pp. 498--502.

\bibitem{panayotov2015librispeech}
V.~Panayotov, G.~Chen, D.~Povey, and S.~Khudanpur, ``Librispeech: an asr corpus
  based on public domain audio books,'' in \emph{2015 IEEE Intl. Conference on
  Acoustics, Speech and Signal Processing (ICASSP)}.\hskip 1em plus 0.5em minus
  0.4em\relax IEEE, 2015, pp. 5206--5210.

\bibitem{boersma2006praat}
P.~Boersma, ``Praat: doing phonetics by computer,''
  \emph{http://www.praat.org/}, 2006.

\bibitem{cho2014learning}
K.~Cho, B.~Van~Merri{\"e}nboer, C.~Gulcehre, D.~Bahdanau, F.~Bougares,
  H.~Schwenk, and Y.~Bengio, ``Learning phrase representations using rnn
  encoder-decoder for statistical machine translation,'' \emph{arXiv preprint
  arXiv:1406.1078}, 2014.

\bibitem{elias2020parallel}
I.~Elias, H.~Zen, J.~Shen, Y.~Zhang, Y.~Jia, R.~Weiss, and Y.~Wu, ``Parallel
  tacotron: Non-autoregressive and controllable tts,'' \emph{arXiv preprint
  arXiv:2010.11439}, 2020.

\bibitem{jordan1999introduction}
M.~I. Jordan, Z.~Ghahramani, T.~S. Jaakkola, and L.~K. Saul, ``An introduction
  to variational methods for graphical models,'' \emph{Machine learning},
  vol.~37, no.~2, pp. 183--233, 1999.

\bibitem{chen2015mxnet}
T.~e.~a. Chen, ``Mxnet: A flexible and efficient machine learning library for
  heterogeneous distributed systems,'' \emph{arXiv preprint arXiv:1512.01274},
  2015.

\bibitem{paleyes2019emulation}
A.~Paleyes, M.~Pullin, M.~Mahsereci, N.~Lawrence, and J.~Gonzalez, ``Emulation
  of physical processes with emukit,'' in \emph{Second Workshop on Machine
  Learning and the Physical Sciences, NeurIPS}, 2019.

\bibitem{guo2020gluoncv}
J.~Guo, H.~He, T.~He, L.~Lausen, M.~Li, H.~Lin, X.~Shi, C.~Wang, J.~Xie, S.~Zha
  \emph{et~al.}, ``Gluoncv and gluonnlp: Deep learning in computer vision and
  natural language processing.'' \emph{Journal of Machine Learning Research},
  vol.~21, no.~23, pp. 1--7, 2020.

\bibitem{michel2011quantitative}
J.-B. Michel, Y.~K. Shen, A.~P. Aiden, A.~Veres, M.~K. Gray, J.~P. Pickett,
  D.~Hoiberg, D.~Clancy, P.~Norvig, J.~Orwant \emph{et~al.}, ``Quantitative
  analysis of culture using millions of digitized books,'' \emph{science}, vol.
  331, no. 6014, pp. 176--182, 2011.

\bibitem{garofolo1993darpa}
J.~S. Garofolo, L.~F. Lamel, W.~M. Fisher, J.~G. Fiscus, and D.~S. Pallett,
  ``Darpa timit acoustic-phonetic continous speech corpus cd-rom. nist speech
  disc 1-1.1,'' \emph{STIN}, vol.~93, p. 27403, 1993.

\bibitem{kominek2004cmu}
J.~Kominek and A.~W. Black, ``The cmu arctic speech databases,'' in \emph{Fifth
  ISCA workshop on speech synthesis}, 2004.

\bibitem{zhao2018l2}
G.~Zhao, S.~Sonsaat, A.~O. Silpachai, I.~Lucic, E.~Chukharev-Hudilainen,
  J.~Levis, and R.~Gutierrez-Osuna, ``L2-arctic: A non-native english speech
  corpus,'' \emph{Perception Sensing Instrumentation Lab}, 2018.

\bibitem{porzuczek2017english}
A.~Porzuczek and A.~Rojczyk, ``English word stress in polish learners speech
  production and metacompetence,'' \emph{Research in Language}, vol.~15, no.~4,
  pp. 313--323, 2017.

\bibitem{oord2018parallel}
A.~Oord, Y.~Li, I.~Babuschkin, K.~Simonyan, O.~Vinyals, K.~Kavukcuoglu,
  G.~Driessche, E.~Lockhart, L.~Cobo, F.~Stimberg \emph{et~al.}, ``Parallel
  wavenet: Fast high-fidelity speech synthesis,'' in \emph{International
  conference on machine learning}.\hskip 1em plus 0.5em minus 0.4em\relax PMLR,
  2018, pp. 3918--3926.

\bibitem{hieke1984linking}
A.~E. Hieke, ``Linking as a marker of fluent speech,'' \emph{Language and
  Speech}, vol.~27, no.~4, pp. 343--354, 1984.

\bibitem{shattuck1994stress}
S.~Shattuck-Hufnagel, M.~Ostendorf, and K.~Ross, ``Stress shift and early pitch
  accent placement in lexical items in american english,'' \emph{Journal of
  Phonetics}, vol.~22, no.~4, pp. 357--388, 1994.

\end{thebibliography}


\end{document}